\documentclass{article}
\usepackage{arxiv}
\usepackage{etoolbox}
\usepackage{booktabs}
\usepackage[hidelinks]{hyperref}
\usepackage{textcomp}
\usepackage{gensymb}
\usepackage[section]{placeins}
\usepackage[nolist]{acronym}
\usepackage[ruled,vlined]{algorithm2e}
\usepackage[long]{optidef}
\usepackage{amssymb}
\usepackage{amsmath}
\usepackage{siunitx}
\usepackage{todonotes}
\usepackage{graphicx}
\usepackage{subfig}

\title{Experimental implementation of an emission-aware prosumer with online flexibility quantification and provision}

\author{
  Hanmin Cai, Philipp Heer \\
  Urban Energy Systems Laboratory \\
  Empa, D\"{u}bendorf, Switzerland \\
  \texttt{hanmin.cai@empa.ch}
}


\usepackage{tikz}
\newcommand*\circled[1]{\tikz[baseline=(char.base)]{
    \node[shape=circle,draw,inner sep=0.5pt] (char) {#1};}}
\begin{document}
\maketitle

\begin{abstract}

Active building energy management holds potential to reduce global energy-related emissions and support flexible operations of future low-carbon systems. This requires to integrate diverse objectives and engage multiple stakeholders. However, there remains a gap in comprehensive field insights into emission reduction, flexibility provision, and user impacts.
This study examined how a real occupied building, with all its energy assets, could function as an emission-aware flexible prosumer. An existing building energy management system was enhanced by integrating a model predictive control strategy. The enhanced setup minimized the equivalent carbon emission due to electricity imports and provided flexibility to the energy system. The experimental results indicated an emission reduction of 12.5\% compared to a rule-based controller that maximized PV self-consumption. In addition, a minimal flexibility provision experiment was demonstrated with a locally emulated distribution system operator. The results suggested that flexibility was provided without the risk of rebound effects. This is due to the flexibility envelope that was self-reported in advance. The study concluded by highlighting technical challenges in realizing emission reduction and flexibility in practice.

\end{abstract}

\acresetall 
\begin{acronym}
\acro{COP}{Coefficient Of Performance}
\acro{DSM}{Demand Side Management}
\acro{DSO}{Distribution System Operator}
\acro{DHW}{Domestic Hot Water}
\acro{EV}{Electric Vehicle}
\acro{HVAC}{Heating, Ventilation and Air Conditioning}
\acro{HP}{Heat Pump}
\acro{ICT}{Information and Communications Technology}
\acro{MPC}{Model Predictive Control}
\acro{OCP}{Optimal Control Problem}
\acro{PV}{Photovoltaics}
\acro{RES}{Renewable Energy Sources}
\acro{RMSE}{Root-Mean-Square Error}
\acro{SH}{Space Heating}
\acro{SOC}{State-Of-Charge}
\acro{TES}{Thermal Energy Storage}

\end{acronym}

\section{Introduction}\label{sec:introduction}
\begin{table*}[htbp]
    \centering
    \label{nomenclature}
    \resizebox{\textwidth}{!}{%
\begin{tabular}{|llllll|}
\hline
\multicolumn{6}{|l|}{Nomenclature}                                                                                       \\
    &                              &      &                                           &     &                            \\
COP & Coefficient of Performance   & HVAC & Heating, Ventilation and Air Conditioning & RES & Renewable Energy Resources \\
DHW & Domestic Hot Water           & HP   & Heat Pump                                 & SH  & Space Heating              \\
DSM & Demand Side Management       & MPC  & Model Predictive Control                  & SOC & State of Charge            \\
DSO & Distribution System Operator & OCP  & Optimal Control Problem                   &     &                            \\
EV  & Electric Vehicle             & PV   & Photovoltaic                              &     &                            \\ \hline
\end{tabular}
    }
\end{table*}
As the share of intermittent \ac{RES} increases and conventional power plants are being phased out, there is an increasing need for flexible demand to maintain security of supply \cite{ostergaard2021energy}. Meanwhile, buildings account for 28\% of the global energy-related CO$_2$ emissions attributed to their operations \cite{global2020}. Hence, active building energy management, falling under the umbrella of \ac{DSM} \cite{palensky2011demand}, can play a critical role in transitioning toward net zero targets. For example, Olivella et al. \cite{olivella2018optimization} show that  actively utilizing buildings' flexibility is a cost-efficient alternative to traditional distribution network expansions when addressing voltage stability issues. 

This new role for buildings introduces diverse objectives and involves a variety of stakeholders. Realistic \ac{DSM} strategies are increasingly needed to consider multiple aspects, namely emission reduction, flexibility quantification and provision. Additionally, occupants' thermal comfort and preferences need to be respected. In fact, the bottom-up nature of building energy flexibility utilization places occupants in the central stage \cite{nagy2023ten,li2022ten}. However, many existing studies examine the topic with limited flexibility resource types, concentrate solely on simulation studies and address just one of the mentioned aspects.

\subsection{Critical literature review}
Bolzoni et al. \cite{bolzoni2021model} develop a \ac{MPC} to manage the flexible assets of a microgrid, reporting considerable emission and cost reductions. 
Although their work includes experimental testing, most control targets are emulated using a Real-Time Digital Simulator (RTDS). Such emulation omits potential disturbances from occupants, which can impact emission reduction and flexibility potentials. Nonetheless, \cite{bolzoni2021model} demonstrates \ac{MPC}'s capability of handling complex tasks.

Reference \cite{heinrich2020ecogrid} provides a retrospective analysis of the flexibility provision in an island-scale experiment, where flexibility is delivered in an open-loop fashion. In addition, Munankarmi et al. \cite{munankarmi2020quantification} examines the flexibility of all behind-the-meter resources of a residential building. This flexibility is further quantified to facilitate interactions between buildings and \acp{DSO}. However, the study relies on perfect knowledge of baseline power and observed rebound effects. 
Reference \cite{de2016quantification} proposes to represent flexibility with a cost curve, comprising sub-optimality gaps due to set-point tracking. 
Maasoumy et al. \cite{maasoumy2014model} propose a scheme that allows utilities to contract buildings for power flexibility provision, and feasible power levels are calculated accounting for operational limits.
Although an experiment is reported, its main purpose is to verify that the high-frequency components of utility's flexibility signal can be tracked. 
Additionally, both \cite{de2016quantification} and \cite{maasoumy2014model} assume that flexibility provision duration is known a priori.

This study focuses on self-reported flexibility, characterized as the capability of modifying energy usage patterns without violating appliance operation limits or compromising end-user preferences and thermal comfort. To this end, the flexibility envelope concept presented in \cite{gasser2021predictive} is adopted. It captures time dependency, the impacts of anticipated weather conditions, and end users' energy usage patterns without assuming knowledge of flexibility provision duration. 
From this perspective, the flexibility quantification in \cite{de2016quantification,maasoumy2014model} can be seen as special cases of the flexibility envelope. However, \cite{gasser2021predictive} only analyzes flexibility quantification in an open-loop fashion and in a pure simulation-based environment. 

\subsection{Unique contribution and paper organization}

Based on the papers reviewed, there remains a gap in comprehensive experimental insights covering three aspects: 1) emission-aware operation, 2) flexibility quantification and provision to \ac{DSO}, and 3) the impacts on/from occupants. The main contributions of this paper are threefold. 
Firstly, we present an emission-aware \ac{MPC} as the base strategy for a prosumer. The emission reduction and occupants' thermal comfort levels are quantified in real-life operation.
Secondly, online flexibility quantification and provision are formulated as optimization problems that can be automated to ease the practical deployment. The framework is experimentally demonstrated with an emulated \ac{DSO} and risk of rebound effects is assessed. Lastly, qualitative and quantitative insights into impacts on/from occupants are reported. While ensuring thermal comfort is incorporated in the control strategy, impacts from occupants are assessed in ex-post analysis.

The remainder of the paper is organized as follows: Section \ref{sec:method} presents the methodological framework of controller design, flexibility quantification and provision. Section \ref{sec:exp_setup} describes the experimental setup and Section \ref{sec:exp_result} presents the results of a week-long experiment, whose real-world implications are critically discussed in Section \ref{sec:discussion}. Finally, Section \ref{sec:conclusion} gives a brief summary and areas for further research are identified.
\section{Methodology}\label{sec:method}
This section first presents control-oriented models and operating constraints of all the assets, followed by the process of parameter identification. Then the \ac{OCP} formulation of the \ac{MPC} controller is provided using the obtained models. Following this, the flexibility envelope calculation and an interaction scheme between the building and the \ac{DSO} are unified with the mentioned \ac{MPC} in one mathematical framework. 

\subsection{Control-oriented model structures for experiments}\label{sec:model_struct}

The considered behind-the-meter assets include \acp{HP} for \ac{SH} and \ac{DHW}, a stationary battery, an \ac{EV} with bidirectional charging, uncontrolled loads and a rooftop \ac{PV} installation. 
The model structures are chosen based on a trade-off among computational complexity, required measurements and interpretability. 

Throughout the paper, we use $\hat{(\cdot)}$ to denote forecast values and use $t$ as the time index. We focus on controlling the active power of behind-the-meter assets while neglecting the reactive power that is in general not available to control for residential loads. 
Power consumption is treated as positive, leading to \ac{PV} power output and battery/\ac{EV} discharging power being negative. Lastly, we use $\mathbb{R}_+$ and $\mathbb{R}_-$ to refer to the set of non-negative and non-positive real values respectively.

\subsubsection{Space heating}
In general, thermal dynamics are highly nonlinear processes. 
However, when the room temperature in buildings is actively controlled within a limited range, the dynamics of each zone can be approximated with an affine model for closed-loop control. Preliminary experimental studies also show that high-order models tend to perform worse than low-order models \cite{antoon2020model}.
The current study considers each room $i \in \mathcal{I}$ as one zone, with the indoor temperature is given by:
\begin{equation}\label{eqn:sh_model}
    T^\text{sh}_{i,t+1} = A^\text{sh} T^\text{sh}_{i,t} + B^\text{sh} \phi^\text{sh}_{i,t} + E^\text{sh} \begin{bmatrix} \hat{T}^\text{amb}_{t} \\ \hat{\phi}^\text{irrd}_{t}\\ 
    \end{bmatrix}, \; \forall t \in \mathcal{H},\; \forall i \in \mathcal{I}
\end{equation}
where $T^\text{sh}_{i,t}$ is the room temperature,  $\phi^\text{sh}_{i,t}$ is the thermal power input, $\hat{T}^\text{amb}_{t}$, $\hat{\phi}^\text{irrd}_{t}$ are the ambient temperature and the solar irradiance forecast respectively, $\mathcal{H}$ is the time horizon, $i$ is the room index and $\mathcal{I}$ is the set of rooms. Since only weather forecast is available in real-time control, $\hat{T}^\text{amb}_{t}$ and $\hat{\phi}^\text{irrd}_{t}$ are used directly.
In addition, we consider continuous thermal power to the rooms and 
the equivalent thermal power output can be obtained by modulating an \ac{HP} operating in an ON/OFF mode with a pulse width modulation strategy. 
The corresponding electrical power consumption of the \ac{HP} is given by:
\begin{equation}\label{eqn:sh_cop}
P^\text{sh}_{t} = \sum_{i\in\mathcal{I}}\phi^\text{sh}_{i,t} / \text{COP}^\text{m}, \; \forall t \in \mathcal{H}
\end{equation}
where COP$^\text{m}$ denotes the COP of the \ac{HP} for \ac{SH}. The corresponding constraints are given by:
\begin{equation}\label{eqn:SH_temp_bound}
T_{i,t}^{\text{sh,min}} - \epsilon_{i,t}^{\text{sh-}} \le T_{i,t}^{\text{sh}} \le T_{i,t}^{\text{sh,max}} + \epsilon_{i,t}^{\text{sh+}}, \; \forall t \in \mathcal{H},\; \forall i \in \mathcal{I}
\end{equation}
\begin{equation}\label{eqn:SH_temp_soft}
\epsilon_{i,t}^{\text{sh-}} \ge 0 ,\; \epsilon_{i,t}^{\text{sh+}} \ge 0, \; \forall t \in \mathcal{H},\; \forall i \in \mathcal{I}
\end{equation}
\begin{equation}\label{eqn:power_up_low_constr_SH}
0 \le P_{t}^{\text{sh}} \le P^{\text{sh,max}}, \; \forall t \in \mathcal{H}
\end{equation}
where $T_{i,t}^{\text{sh,max}}$ and $T_{i,t}^{\text{sh,min}}$ are the predefined time-varying upper and lower indoor temperature limits respectively, $\epsilon_{i,t}^{\text{sh-}}$ and $\epsilon_{i,t}^{\text{sh+}}$ are the slack variables introducing soft constraints to ensure feasible solutions, and $P^{\text{sh,max}}$ is the electrical power capacity of the \ac{HP}. 

\subsubsection{Domestic hot water heating}
\ac{DHW} is supplied by a fixed-speed \ac{HP} operating on ON/OFF mode with a buffer tank. Stratification effects within the tank are neglected, and the average tank temperature is given by:
\begin{equation}\label{eqn:dhw_model}
    T^\text{dhw,avg}_{t+1} = A^\text{dhw} T^\text{dhw,avg}_{t} + B^\text{dhw} \phi^\text{dhw}_{t} + E^\text{dhw}  m^\text{draw}_{t}, \; \forall t \in \mathcal{H}
\end{equation}
where $T^\text{dhw, avg}_{t}$ is the volume-weighted average tank temperature, $\phi^\text{dhw}_{t}$ is the thermal power input and $m^\text{draw}_{t}$ is the amount of water draw.
The corresponding electrical power consumption of the \ac{HP} can be given by:
\begin{equation}\label{eqn:dhw_cop}
P^\text{dhw}_t = \phi^\text{dhw}_t / \text{COP}^\text{h}, \; \forall t \in \mathcal{H}
\end{equation}
where $\text{COP}^\text{h}$ is used to differentiate from the above-mentioned \ac{SH} \ac{HP}. 
This distinction is made because \ac{SH} and \ac{DHW} typically have different inlet temperatures. Average tank temperature constraints are given by:
\begin{equation}\label{eqn:DHW_temp_bound}
T_t^{\text{dhw,min}} - \epsilon_{t}^{\text{dhw-}} \le T_{t}^{\text{dhw}} \le T_t^{\text{dhw,max}} + \epsilon_{t}^{\text{dhw+}}, \; \forall t \in \mathcal{H}
\end{equation}
\begin{equation}\label{eqn:DHW_temp_soft}
\epsilon_{t}^{\text{dhw-}} \ge 0 ,\; \epsilon_{t}^{\text{dhw+}} \ge 0, \; \forall t \in \mathcal{H}
\end{equation}
where $T_t^{\text{dhw,max}}$ and $T_t^{\text{dhw,min}}$ are the predefined time-varying upper and lower average tank temperature limits respectively, $\epsilon_{t}^{\text{dhw-}}$ and $\epsilon_{t}^{\text{dhw+}}$ are the slack variables enforcing soft constraints. The ON/OFF operating mode is modeled with a binary variable $z_{t}^{\text{dhw}}$:
\begin{equation}\label{eqn:power_up_low_constr_DHW}
P_{t}^{\text{dhw}} = P^{\text{dhw,max}}z_{t}^{\text{dhw}}, \; \forall t \in \mathcal{H}
\end{equation}
where $P^{\text{dhw,max}}$ is the electric power capacity of the \ac{DHW} \acp{HP}. 

\subsubsection{Stationary electric battery}
A model that captures battery self-losses, charging/discharging efficiencies is given by:
\begin{equation}\label{eqn:ebat_model}
    \text{SOC}_{t+1}^{\text{ebat}} = A^\text{ebat}\text{SOC}_{t}^{\text{ebat}} + B^\text{ebat} \begin{bmatrix} P_{t}^{\text{ebat, ds}} \\ P_{t}^{\text{ebat, ch}} \\ 
    \end{bmatrix}, \; \forall t \in \mathcal{H}
\end{equation}
where $\text{SOC}_{t}^{\text{ebat}}$ is the battery \ac{SOC}, $P_{t}^{\text{ebat, ch}} \in \mathbb{R}_+$ and $P_{t}^{\text{ebat, ds}} \in \mathbb{R}_-$ are the battery charging and discharging power respectively. The mutual exclusiveness of $P_{t}^{\text{ebat, ch}}$ and $P_{t}^{\text{ebat, ds}}$ is enforced with a binary variable $z_{t}^{\text{ebat}}$ as follows.
\begin{equation}\label{eqn:bat_ch_power}
0 \le P_{t}^{\text{ebat, ch}} \le P^{\text{ebat,max}} z_{t}^{\text{ebat}}, \; \forall t \in \mathcal{H}
\end{equation}
\begin{equation}\label{eqn:bat_ds_power}
-P^{\text{ebat,max}} (1-z_{t}^{\text{ebat}}) \le P_{t}^{\text{ebat, ds}} \le 0, \; \forall t \in \mathcal{H}
\end{equation}
\begin{equation}\label{eqn:ebat_soc_bound}
\text{SOC}^{\text{ebat,min}} - \epsilon_{t}^{\text{ebat}} \le \text{SOC}_{t}^{\text{ebat}} \le 100, \; \forall t \in \mathcal{H}
\end{equation}
\begin{equation}\label{eqn:ebat_soft}
\epsilon_{t}^{\text{ebat}} \ge 0, \; \forall t \in \mathcal{H} 
\end{equation}
where $\epsilon_{t}^{\text{ebat}}$ is the slack variables introducing soft constraints to ensure feasible solutions and $\text{SOC}^{\text{ebat,min}}$ is the minimum \ac{SOC}.

\subsubsection{Electric vehicle}
Similar to the stationary battery model, the \ac{EV} battery is modeled as follows:
\begin{equation}\label{eqn:ev_model}
    \text{SOC}_{t+1}^{\text{ev}} = A^\text{ev}\text{SOC}_{t}^{\text{ev}} + B^\text{ev} \begin{bmatrix} P_{t}^{\text{ev, ds}} \\ P_{t}^{\text{ev, ch}} \\ 
    \end{bmatrix}, \; \forall t \in \mathcal{H}
\end{equation}
where $\text{SOC}_{t}^{\text{ev}}$ is the \ac{EV} battery \ac{SOC}, $P_{t}^{\text{ev, ch}} \in \mathbb{R}_+$ and $P_{t}^{\text{ev, ds}} \in \mathbb{R}_-$ are the battery charging and discharging power respectively. 
The limits on \ac{SOC} is given by:
\begin{equation}\label{eqn:EV_soc_bound}
\text{SOC}^{\text{ev,min}}  - \epsilon_{t}^{\text{ev}} \le \text{SOC}_{t}^{\text{ev}} \le 100, \; \forall t \in \mathcal{H}
\end{equation}
\begin{equation}\label{eqn:ev_ebat_soft}
\epsilon_{t}^{\text{ev}} \ge 0 ,\; \forall t \in \mathcal{H} 
\end{equation}
where $\epsilon_{t}^{\text{ev}}$ is the slack variable introducing soft constraints and $\text{SOC}_{t}^{\text{ev,min}}$ is the minimum \ac{SOC} of \ac{EV} battery. Simultaneous charging/discharging are avoided by introducing a binary variable $ z_{t}^{\text{ev}}$ as follows.
\begin{equation}\label{eqn:ev_ch_power}
0 \le P_{t}^{\text{ev, ch}} \le P^{\text{ev,max}} z_{t}^{\text{ev}}, \; \forall t \in \mathcal{H}
\end{equation}
\begin{equation}\label{eqn:ev_ds_power}
-P^{\text{ev,max}} (1-z_{t}^{\text{ev}}) \le P_{t}^{\text{ev, ds}} \le 0, \; \forall t \in \mathcal{C}
\end{equation}
\begin{equation}\label{eqn:ev_notpresent_power}
P_{t}^{\text{ev, ch}} = 0, \; P_{t}^{\text{ev, ds}} = 0, \; \forall t \in \mathcal{H} \backslash \mathcal{C}
\end{equation}
\begin{equation}\label{eqn:EV_soc_bound_dep}
\text{SOC}^{\text{ev},\text{min}}_{\text{sup } \mathcal{C}} \le \text{SOC}_{\text{sup } \mathcal{C}}^{\text{ev}} \le 100
\end{equation}
where $\mathcal{C}$ is the set of time steps when the \ac{EV} is connected to the charger at home. 
Constraint \eqref{eqn:ev_notpresent_power} indicates \ac{EV} is not available for control when absent. Constraint \eqref{eqn:EV_soc_bound_dep} ensures minimum departure state-of-charge $\text{SOC}^{\text{ev},\text{min}}_{\text{sup } \mathcal{C}}$ with $\text{sup } \mathcal{C}$ being the departure time instant or supremum of the set $\mathcal{C}$.

\subsubsection{Photovoltaic} \label{sec:pv_model}
\ac{PV} power output is predicted based on weather forecast and the formulation is given by:
\begin{equation}\label{eqn:pv}
    \hat{P}_t^\text{pv} = \beta_0^\text{pv} + \beta_1^\text{pv} \hat{\phi}^\text{irrd}_t + \beta_2^\text{pv} \hat{T}_t^\text{amb} + \epsilon
\end{equation}
where $\hat{P}_{t}^{\text{pv}}\in \mathbb{R}_-$ is the predicted \ac{PV} output. Such numerical weather condition-based forecasting is effective for horizons exceeding 4 hours \cite{ahmed2020review}, whereas persistence forecast is popular for short-term PV forecasting.
Hence, \ac{PV} power forecast $\{\hat{P}_{t}^{\text{pv}}|t\in \mathcal{H}\}$ within the prediction horizon $\mathcal{H}$ is additionally combined with a persistence forecast. Since PV output forecasting is not the focus of this work, the combination of both forecast is determined empirically.
The result is used in the predictive controller.  
Coefficients ($\beta_0^\text{pv}, \beta_1^\text{pv}, \beta_2^\text{pv}$) are re-identified every day accounting for the impact of potential coverings on the installation, such as leaves and snow. 
The coefficients are updated if the re-identification is satisfactory.

\subsubsection{Uncontrolled loads}
Uncontrolled loads refer to lighting, cooking and wet appliances. This study aggregates the power $\hat{P}_t^\text{fix}\in \mathbb{R}_+$ of these appliances to account for their electricity consumption. The forecast is a combination of a 15-minute ahead persistence forecast and a 24-hour ahead persistence forecast.

\subsubsection{Entire building}
The energy balance of the entire building is given by:
\begin{equation}\label{eqn:SH_plus_DHW_power}
P_{t}^\text{btg} + P_{t}^\text{gtb} =  P_{t}^{\text{sh}}+P_{t}^{\text{dhw}}+P_{t}^{\text{ebat,ch}} +P_{t}^{\text{ebat,ds}}+ P_{t}^{\text{ev,ch}} + P_{t}^{\text{ev,ds}}+\hat{P}_{t}^{\text{pv}}+ \hat{P}_t^\text{fix},\; \forall t  \in \mathcal{H} 
\end{equation}
where $P_{t}^\text{btg} \in \mathbb{R}_-$ and $P_{t}^\text{gtb} \in \mathbb{R}_+$ are the electricity export and import respectively. They are differentiated because the total equivalent carbon emission is calculated according to the carbon intensity of electricity imported from the grid, and it is assumed that electricity export does not offset building's carbon footprint.
To ensure the mutual exclusiveness of $P_{t}^\text{btg}$ and $P_{t}^\text{gtb}$, a binary variable $z_{t}^{\text{grid}}$ is introduced.
\begin{equation}\label{eqn:net_exchange_max}
0 \le P_{t}^{\text{gtb}} \le M z_{t}^{\text{grid}}, \; \forall t \in \mathcal{H}
\end{equation}
\begin{equation}\label{eqn:net_exchange_min}
-M (1-z_{t}^{\text{grid}}) \le P_{t}^{\text{btg}} \le 0, \; \forall t \in \mathcal{H}
\end{equation}
where $M \in \mathbb{R}_+$ is a sufficiently large constant.

\subsection{Parameter identification}\label{sec:model_struct}

A prediction error method is used, which can be described as minimizing the differences between the observed
outputs and the predicted outputs based on the model and the historical data \cite{ljung1998system}. For conciseness, a formulation is given as follows:
\begin{equation}\label{eqn:system_model}
     y_{t} = \hat{A}y_{t-1} + \hat{B}u_{t-1} + \epsilon_{t}
\end{equation}
where $\hat{A}$, $\hat{B}$ denote the system dynamics-related parameters that need to be identified. Eq. \eqref{eqn:system_model} assumes a state-space representation with an identity matrix as its output matrix and a zero matrix as its feed-through matrix.
The $k$-step ahead prediction based on the historical data and the model structure is formulated as follows:
\begin{equation}\label{eqn:model_assumed}
    \hat{y}_{t} = \hat{A}^ky_{t-k} + \sum_{i=1}^k \hat{A}^{k-i}\hat{B}u_{t-k-1+i}
\end{equation}
where $\hat{y}_t$ is the $k$-step ahead prediction based on the historical inputs and the boundary conditions described by the set $\{u_\iota|\iota\in [t-k, t]\}$. 
The unknown parameters are obtained by optimizing the cost function consisting of accumulated prediction errors as follows:
\begin{equation}\label{eqn:model_id}
   J = \sum_{t=1}^N \frac{1}{N}L_{\delta }(\hat{y}_{t},y_{t})
\end{equation}
where $N$ denotes the number of prediction error terms and $y_t$ is the observed output. Notably, the Huber function $L_{\delta }(\cdot)$ to compute losses instead of a square function to reduce the impacts of outliers. The optimal parameters $\hat{A}^*$ and $\hat{B}^*$ minimizes the cost $J$. For this study, a prediction horizon of 24 hours is chosen.

\subsection{Optimal control problem formulation}\label{sec:mpc_formulation}
The full \ac{OCP} formulation of the \ac{MPC} is given by:
\begin{mini!}[2]
	{\{\mathbf{X}_t,\mathbf{\mathbf{U}}_t,\mathbf{V}_t|\forall t \in \mathcal{H}\}}
	{\sum_{t \in \mathcal{H}}c^\text{emi}_{t}P_{t}^\text{gtb}\Delta \tau_t + L_t\Delta \tau_t \label{eqn:cost_function}}   
	{\label{eqn:gb1_optimization}}             
	{}                              
	\addConstraint{\text{Eq. } \eqref{eqn:sh_model} \text{ to Eq. } \eqref{eqn:net_exchange_min} \label{eqn:first_cons_full}}
\end{mini!}
where $L_t :=  w_1(L_{\delta }(\epsilon_{i,t}^{\text{sh+}}) + L_{\delta }(\epsilon_{t}^{\text{dhw+}})) + w_2(L_{\delta}(\epsilon_{i,t}^{\text{sh-}}) + L_{\delta}(\epsilon_{t}^{\text{dhw-}})) + w_3 ((\epsilon_{t}^{\text{ev}})^2 + (\epsilon_{t}^{\text{ebat}})^2)$ denotes the cost associated with soft constraints, Huber function $L_{\delta}(\cdot)$ is used to formulate the penalties of constraint violations instead of quadratic penalties to be robust to outliers.
, $ w_1 $, $ w_2 $ and $ w_3 $ are the customized weighting factors, $\{\mathbf{X}_t|\forall t \in \mathcal{H}\} :=\{T_{i,t}^{\text{sh}},$ $T_{t}^{\text{dhw}},$ $\text{SOC}_{t}^{\text{ev}},$ $\text{SOC}_{t}^{\text{ebat}}|\forall t \in \mathcal{H},\;\forall i \in \mathcal{I}\}$ is the set of state variables, $\{\mathbf{U}_t|\forall t \in \mathcal{H}\} := \{P_{t}^{\text{sh}},$ $ P_{t}^{\text{dhw}},$ $ P_{t}^{\text{ebat,ch}},$ $ P_{t}^{\text{ebat,ds}},$ $  P_{t}^{\text{ev,ch}},$ $  P_{t}^{\text{ev,ds}}|\forall t \in \mathcal{H}\}$ is the set of control input variables, $\{\mathbf{V}_t|\forall t \in \mathcal{H}\} := \{\epsilon_{i,t}^{\text{sh-}},$ $ \epsilon_{i,t}^{\text{sh+}},$ $ \epsilon_{t}^{\text{dhw-}},$ $ \epsilon_{t}^{\text{dhw+}},$ $ \epsilon_{t}^{\text{ev}},$ $ \epsilon_{t}^{\text{ebat}},$ $ z_{t}^{\text{ebat}},$ $ z_{t}^{\text{ev}},$ $ z_{t}^{\text{grid}},$ $ z_{t}^{\text{dhw}}, P_{t}^\text{btg},$ $ P_{t}^\text{gtb}|\forall t \in \mathcal{H},\;\forall i \in \mathcal{I}\}$ denotes the set of the remaining decision variables, $c^\text{emi}_{t} \in \mathbb{R}_+$ is the grid carbon intensity.
It needs to be mentioned that only the variables in $\{\mathbf{U}_t|\forall t \in \mathcal{H}\}$ can be physically controlled. Although the rest of the variables are formulated as decision variables of the \ac{OCP}, they can be determined by $\{\mathbf{U}_t|\forall t \in \mathcal{H}\}$, system dynamics equations, and algebraic constraints. 

\begin{figure*}[!htbp]
	\centering
    \includegraphics[width=0.8\textwidth]{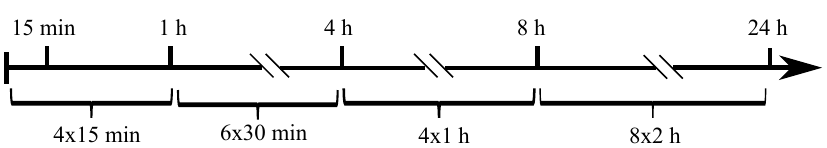}
	\captionof{figure}{Inequidistant sampling time  $\Delta \tau$ over an optimization horizon of 24 hours.}\label{fig:dyna_sampling_time}
\end{figure*}

In addition, a set of inequidistant sampling time is used to reduce the number of decision variables as shown in \autoref{fig:dyna_sampling_time}. Larger sampling time  are assumed for the time instants further in the optimization horizon, as forecast and modeling errors increase as well. The horizon is chosen to be 24 hours to include future knowledge such as \ac{PV} power output and \ac{EV} usage patterns.

\subsection{Flexibility envelope}\label{sec:flex_env_form}
The flexibility envelope, introduced qualitatively in a previous study \cite{gasser2021predictive},  is formulated as optimization problems and extended to consider the flexibility from batteries, \acp{EV} with bidirectional charging, and curtailable PV.

\begin{figure}[!htbp]
	\centering
	\includegraphics[width=.8\textwidth]{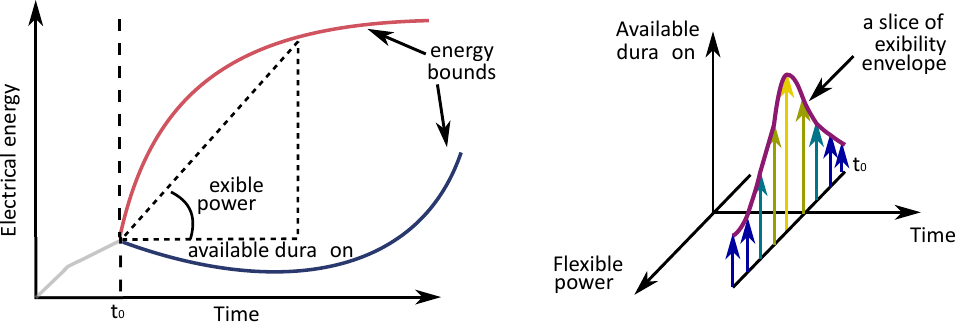}
	\caption{Illustration of the workflow to obtain one slice of a flexibility envelope at $t_0$. The left figure depicts upper and lower energy bounds derived from extreme scenarios. The bounds indicate flexible power and the corresponding available duration. The right figure maps the power levels and duration onto a three-dimensional space and illustrates one slice of flexibility envelope. All slices at each time step within the horizon constitute a full flexibility envelope.}
	\label{fig:flex_envelope}
\end{figure}

The flexibility envelope identification starts with identifying the energy bounds by energizing flexible appliances to their extremes. The upper energy bound is identified by maximizing device consumption as early as possible, coupled with full \ac{PV} output curtailment.
To identify the lower energy bound, all the loads are set to consume as late and as little as possible. Simultaneously, the stationary and \ac{EV} batteries are set to discharge as early and as much as possible without curtailment of \ac{PV} output.
The upper and lower energy bounds are illustrated as the red and blue curves in \autoref{fig:flex_envelope}. 
Formally, the above-mentioned notion can be formulated as optimization problems by modifying the \ac{OCP} formulation in Eq. \eqref{eqn:gb1_optimization}. Initially, solving the original \ac{OCP} gives $\{\Tilde{\mathbf{X}}_t, \Tilde{\mathbf{U}}_t, \Tilde{\mathbf{V}}_t|\forall t\in \mathcal{H}\}$, in which $\{\Tilde{\mathbf{X}}_t|\forall t\in \mathcal{H}\}$ are the optimal state trajectories. 
The cost functions for deriving the upper and the lower energy bounds are defined as $J^\uparrow := \sum_{t\in \mathcal{H}} L_t - e^{-\frac{t}{\rho}}\mathbf{U}_t$
and
$J^\downarrow := \sum_{t \in \mathcal{H}}L_t + e^{-\frac{t}{\rho}}\mathbf{U}_t $ respectively, in which $e^{-\frac{t}{\delta}}$ is an exponentially decaying weighting factor.
Moreover, these optimization problems are initialized using the optimal state trajectories $\Tilde{\mathbf{X}}_t$. 
Solving the \ac{OCP} with the new cost functions, we have $\{\mathbf{U}^\uparrow_t|\forall t\in \mathcal{H}'\} = \text{argmin}_{\{\mathbf{X}_t, \mathbf{U}_t, \mathbf{V}_t|\forall t \in \mathcal{H}'\}}J^\uparrow$
and $\{\mathbf{U}^\downarrow_t|\forall t\in \mathcal{H}'\} = \text{argmin}_{\{\mathbf{X}_t, \mathbf{U}_t, \mathbf{V}_t|\forall t \in \mathcal{H}'\}}J^\downarrow$, in which $\mathcal{H}'$ is the optimization horizon for energy bounds identification and it might be different from $\mathcal{H}$.
Denote the aggregate power of all assets except uncontrolled loads as $P_t\in \mathbb{R}$ and we further have $P^\uparrow_t$ and  $P^\downarrow_t$ as the aggregated power calculated from $\mathbf{U}^\uparrow_t$ and $\mathbf{U}^\downarrow_t$ respectively. With $\{P^\downarrow_t, P^\uparrow_t|\forall t\in \mathcal{H}'\}$, the upper and lower energy bounds can be obtained as $\{E^\uparrow_n := \sum_{k=1}^nP^\uparrow_k\Delta\tau_k|\forall n\in \mathcal{H}'\}$ and $\{E^\downarrow_n := \sum_{k=1}^nP^\downarrow_k\Delta\tau_k|\forall n\in \mathcal{H}'\}$. 
The energy bounds of the PV installation are calculated differently as there is no inter-temporal correlation. For a curtailable PV, the upper and the lower energy bounds are given by $\{E^\uparrow_n := 0|\forall n\in \mathcal{H}'\}$ and $\{E^\downarrow_n := \sum_{k=1}^n\hat{P}^\text{pv}_k\Delta\tau_k|\forall n\in \mathcal{H}'\}$ respectively. Finally, we can obtain the energy bounds of the entire building combining the above-mentioned bounds. 
To ease the computational efforts, power of all assets is approximated as continuous variables, at the example of introducing biases. 

In essence, system dynamics and operational limits dictate the energy bounds, which encapsulate all feasible energy trajectories.
$\{P_k|E^\downarrow_n \le \sum_{k=1}^nP_k\Delta\tau_k \le E^\uparrow_n, n\in \mathcal{H}',k\in \mathcal{H}'\}$, which is an infinite set due to continuously controllable set points of the components. 
To ease representation, we consider only power trajectories with invariant power levels as the dashed line in \autoref{fig:flex_envelope} shows with the slope indicating the power. The corresponding available duration is limited by the second endpoint of the line.
Denote a future time instant as $\tau \in \mathcal{T}_+$ and feasible power level as $P\in \mathcal{P}_\tau$. The mentioned available duration is represented as an implicit function of power level and time $f: \mathbb{R} \times  \mathcal{T}_+ \xrightarrow{}  \mathbb{R}_+$. 
To summarize, the flexibility envelope is a three-dimensional surface, comprised of points in the set \{($\tau$, $P$, $f(\tau, P)$)$| \tau \in  \mathcal{T}_+,P\in \mathcal{P}_\tau$\}. 

\subsection{Interaction between buildings and a \ac{DSO}}
The proposed flexibility envelope captures the thermal inertia of a building, the storage of a domestic hot water tank, the bidirectional charging of a stationary electric battery and/or an \ac{EV}, and the curtailable \ac{PV} power. When this envelope is self-reported in advance, a \ac{DSO} obtains a comprehensive overview of the available flexibility at a given building. 
Upon receiving the flexibility envelope, the \ac{DSO} sends a flexibility provision message $(\tau_\text{s},\tau_\text{e}, P)$ to the building, where $\tau_\text{s}$ and $\tau_\text{e}$ denote the starting time and the ending time of flexibility provision respectively, and $P\in \mathcal{P}_\tau$ denotes the power level that needs to be tracked. By definition, there is $\tau_\text{e} - \tau_\text{s} \le f(\tau, P)$. Upon receiving $(\tau_\text{s},\tau_\text{e}, P)$, the cost function of the \ac{OCP} is extended. It now includes the cost of tracking errors with the weighting factor $\omega_4$ and the resultant is $J:=\sum_{t \in \mathcal{H}}c^\text{emi}_{t}P_{t}^\text{gtb}\Delta \tau_t + L_t\Delta \tau_t+\omega_4(P_t^\text{gtb}+P_t^\text{btg}-\hat{P}_t^\text{fix}-P)^2\Delta\tau_t$.

\begin{figure}[!htbp]
	\centering
	\includegraphics[width=.45\textwidth]{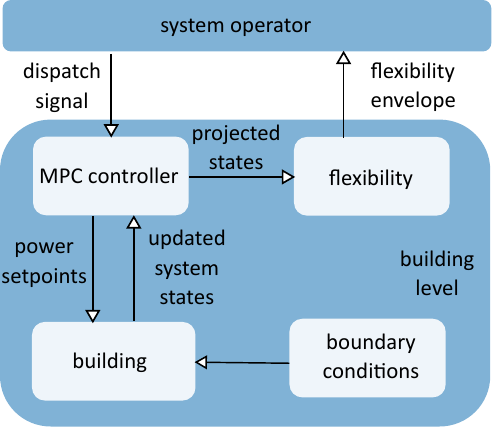}
	\caption{Illustration of the information flow among the building, the controller and the system operator. At each time step, updated building state measurements are retrieved by the controller to solve the emission-aware \ac{OCP}. The resultant state trajectories are used to quantify the flexibility envelope, which is self-reported to the \ac{DSO}. Upon receipt, the \ac{DSO} may request the prosumer to provide flexibility depending on the condition of the network.}\label{fig:overall_struct}
\end{figure}

Moreover, the overall two-stage framework is illustrated in \autoref{fig:overall_struct}. This framework allows the \ac{DSO} to address local network issues using local flexibility resources. When there is a sufficient number of buildings supporting this framework, the \ac{DSO} can even adjust the load to follow the expected supply of the system. 
An implementation example is given in Section \ref{sec:exp_setup} to complement the conceptual schematic.

\section{Experimental case study}\label{sec:exp_setup}
\subsection{Hardware specification}
\begin{figure*}[!htbp]
	\centering
    \includegraphics[width=\textwidth]{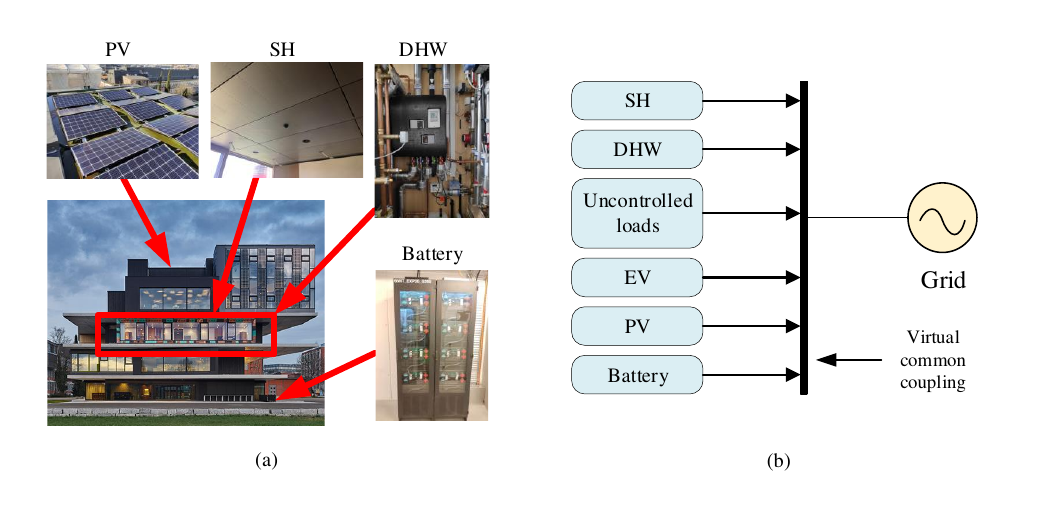}
	\captionof{figure}{Physical layer of the experimental setup (a) and its single-line representation (b). The physical layer shown in (a) includes \ac{SH} with ceiling heating panels, \ac{DHW} with a buffer tank, fixed loads of an apartment unit marked in the red box, a PV installation and a battery. The \ac{PV} installation is placed on the roof and the battery is located in the basement, which are not directly visible in the figure. The bi-directional charging \ac{EV} is a simulated entity and is not visible here. All pictures are taken from \cite{nestpubwiki}. The spatially distributed hardware are virtually coupled via their timestamped measurements as shown in (b). Such a virtual coupling is seen as a billing point for the \ac{DSO}.}\label{fig:nest}
\end{figure*}

All assets, except the \ac{EV}, were real components in the experiment. The \ac{EV} with bidirectional charging was simulated with an identical model for both the emulator and the controller. This assumes perfect modeling of the \ac{EV} battery and charging/discharging process.
Physical components are from the NEST demonstrator at Empa in Switzerland \cite{richner2018nest}, as \autoref{fig:nest} shows. The hardware was distributed around the research infrastructure. The time-stamped measurements allowed emulating the actual operation of a prosumer equipped with all the aforementioned assets.
The total power of the emulated building combined all power measurements and the simulated \ac{EV} power.

The \ac{SH} and \ac{DHW} under control came from the residential unit marked with a red box in \autoref{fig:nest}. 
The UMAR unit is equipped with water-based ceiling panels for space heating. 
The room temperature comfort zone of the whole unit was set as [\SI{22}{\degreeCelsius}, \SI{23}{\degreeCelsius}]. The range was relaxed to [\SI{21}{\degreeCelsius}, \SI{25}{\degreeCelsius}] during the daytime, specifically between $8$~am and $8$~pm, as occupants are likely absent during these hours. Occupants can adjust the comfort zone without impacting the method's generality.

For domestic hot water heating, the tank's average temperature was maintained between [\SI{45}{\degreeCelsius}, \SI{60}{\degreeCelsius}]. The lower limit was boosted to [\SI{59}{\degreeCelsius}, \SI{60}{\degreeCelsius}] at least once a week to avoid Legionella contamination \cite{cai2018technical}. In our experiment, this timing was manually chosen to be Sunday morning between $4$~am and $6$~am. Regarding future water draw, a persistence forecast was used assuming that the future water draw will be the same as the day one week earlier. Preliminary studies employed a recurrent neural network and a Markov chains-based approach with one-year data. However, these did not enhance the forecast accuracy. This was due to the highly stochastic nature of human behaviour at the household level. 
Uncontrolled loads such as the cooking stove and the dishwasher were also located at the same unit. Therefore, consistent occupancy patterns for all the assets were observed.

A Lithium-ion battery of 17.5~kWh with bi-directional charging/discharging power of $\pm$5~kW was operated\footnote{The onsite battery has a capacity of 96 kWh and was artificially limited its operating range to emulate a residential battery system.}. Conversely, an \ac{EV} of 50~kWh with bi-directional charging/discharging power of $\pm$7~kW was included as a simulated system component in the experiment. \ac{EV} arrival and departure times were assumed to be known due to driver's input, which aligned with the large-scale field trial in \cite{dudek2021flexibility}. \ac{PV} electricity output was considered emission-free, but the power exported to the grid did not offset the building's carbon footprint. Interested readers are referred to \cite{exp_specs} for comprehensive description of test specifications and experiment plans.

\subsection{Communication architecture}
\begin{figure*}[!htbp]
	\centering
    \includegraphics[width=.8\textwidth]{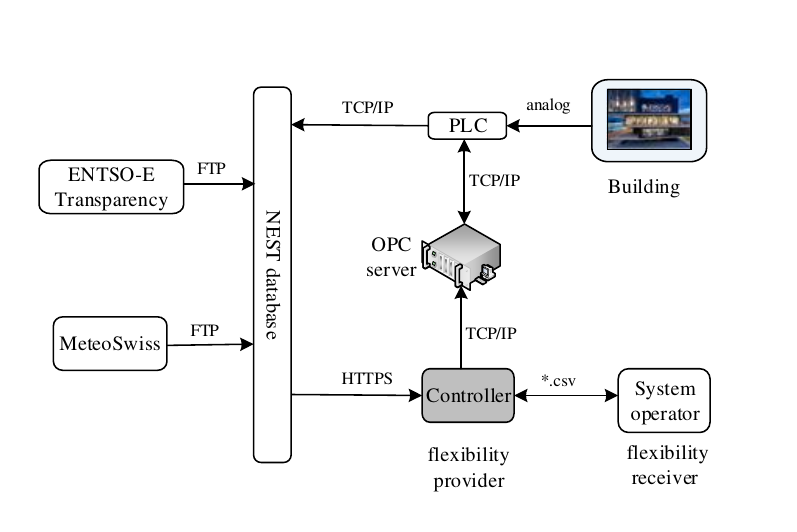}
	\captionof{figure}{Information exchange among the controller, the external information sources (e.g., MeteoSwiss), an emulated system operator and the hardware in the experiments.}\label{fig:communication_architecture}
\end{figure*}

The interactions in the proposed system are illustrated in \autoref{fig:communication_architecture}.
Measurements were obtained  from the Microsoft SQL database every 15~minutes \cite{nestpubwiki}. The forecast for ambient temperature and global solar irradiation was provided by the Federal Office of Meteorology and Climatology (MeteoSwiss) \cite{steppeler2003meso} every~12 hours.
Actuation setpoints were communicated through TwinCAT PLCs, which interacted with the hardware using ModBus or analog signals. The controller and PLCs are connected via an OPC server \cite{opcua}. For a detailed description, see \cite{exp_specs}. The electricity carbon intensity data for Switzerland came from external sources \cite{aliunid_proj_report}, using data from the ENTSO-E Transparency Platform \cite{entsoe}. Communication between the controller and the emulated \ac{DSO} was achieved using local csv files.  For more details, see \cite{exp_specs} and \cite{aliunid_proj_report}.

\section{Results and analyses}\label{sec:exp_result}

This section delves into closed-loop control results, the controller's performance, and flexibility quantification and provision.

\subsection{Closed-loop control results}
The \ac{MPC} controller solved the \ac{OCP} from Section \ref{sec:mpc_formulation} every 15~minutes. The \ac{OCP} was formulated with CVXPY \cite{diamond2016cvxpy} in Python 3 and solved using MOSEK \cite{mosek}.The control variables' optimal values were obtained over a 24h horizon. Only the next time step's decision values are relayed to the actuators via the Python-OPC UA client. All the hardware is equipped with dedicated meters, and all the measurements are stored in a Microsoft SQL database \cite{nestpubwiki}. Since the actuators cannot perfectly execute the set points, a distinction is made between the measured power and the power set point planned by the controller in the following results.

Model parameters were identified from 2019 data using the prediction error method \cite{ljung1987systems}. The models are then re-sampled according to the set of inequidistant sampling time intervals shown in \autoref{fig:dyna_sampling_time}. The strategy assumes the same power set point during a sample time interval. The accuracy of each individual appliance is not presented here. However, additional experiments were carried out, where energy input for \ac{SH} and \ac{DHW} was minimized and their temperature were found to stay close to their lower limits \cite{aliunid_proj_report}. It is worth noting that this is a common experimental design for validating modeling accuracy, as demonstrated in \cite{bunning2020experimental} and \cite{lian2021adaptive}. Additionally, the following experimental results further confirm modeling accuracy, as both \ac{SH} and \ac{DHW} temperatures stay within predefined limits while optimizing according to carbon intensity, except for dramatic disturbances.

In the week-long experiment, power set points of all controllable assets were obtained by solving the \ac{OCP} formulated in Eq. \eqref{eqn:gb1_optimization} and the \ac{OCP} with a modified objective function during flexibility provision. This section presents the controller decisions, realized power input and responses of all the assets, first detailing results for individual appliance.  The net power exchange with the grid follows, as illustrated in \autoref{fig:exp_result_grid_emission_nov}.

\subsubsection{Space heating}
\begin{figure*}[!htp]
	\centering
    \includegraphics[width=.85\textwidth]{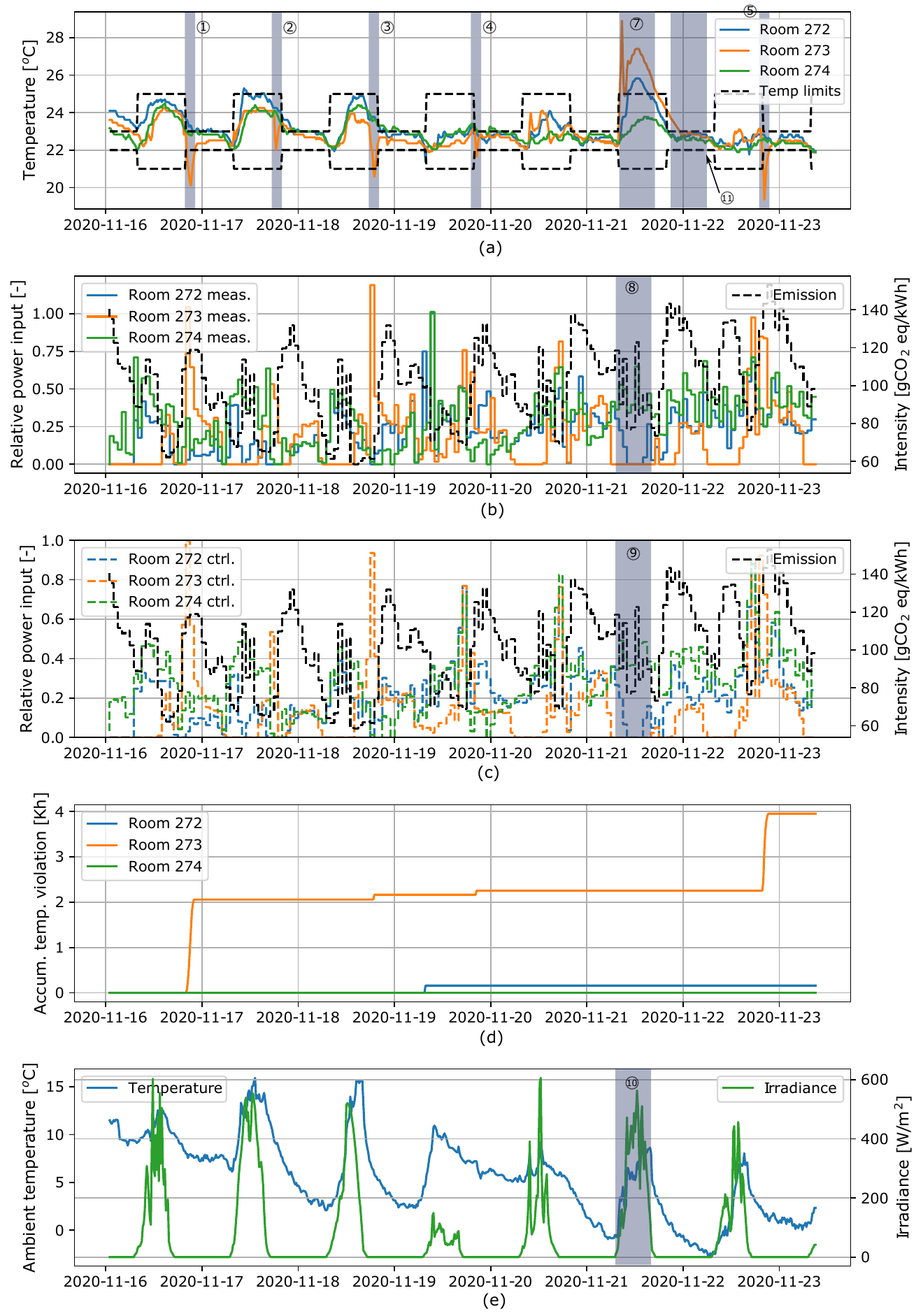}
	\captionof{figure}{Field experiment results of \ac{SH}. (a) shows the measured temperatures of all three rooms. (b) shows the measured thermal power input into each room. (c) shows the thermal power input planned by the controller. The dashed black lines in (b) and (c) indicate the carbon-intensity profile. (d) shows the time-integral of room temperature deviations below the lower limit, measured in kelvin hours (Kh). (e) shows the ambient temperature and the irradiance measurements.}\label{fig:exp_result_sh_emission_nov}
\end{figure*}

The temperatures of all three rooms are shown to approach their upper temperature limits in \autoref{fig:exp_result_sh_emission_nov}(a). This reflects the controller's strategy: maximizing electricity use when carbon intensity was low, and minimizing it when high. Significant temperature drops in Room 273 (marked by the grey periods \circled{1} - \circled{5} in the figure) were due to extend window openings for hygiene, related to Covid. Excluding these instances, indoor temperature generally remained within the comfort zone. 

Comparison between \autoref{fig:exp_result_sh_emission_nov}(b) and \autoref{fig:exp_result_sh_emission_nov}(c) indicate that actuators mostly follow the controller decisions even though there are mismatches. 
Figure \ref{fig:exp_result_sh_emission_nov}(d) shows that  temperatures within Room 272 and Room 274 are mostly within the limits, whereas the accumulated temperature deviation reached around 4~Kh in Room 273 by the end of the experiment due to dramatic disturbances.
We can also observe from \autoref{fig:exp_result_sh_emission_nov}(a) that indoor temperature occasionally exceeds the upper temperature limit such as the grey period \circled{7}. In general, such behaviour can be attributed to factors such as high solar irradiance, modeling errors, and internal gains that are not entirely captured. Notably, the solar irradiance as shown in the grey period \circled{10} in \autoref{fig:exp_result_sh_emission_nov}(e) is not substantially higher than other days. Additionally, the measurements of the grey period \circled{8} in \autoref{fig:exp_result_sh_emission_nov}(b) and the grey period \circled{9} in \autoref{fig:exp_result_sh_emission_nov}(c) show that the controller decision and actual power input to Room 272 and Room 273 were close to 0~kW during the same period. Hence, the overshoot is more likely caused by dramatic internal gains.

\subsubsection{Domestic hot water}
\begin{figure}[!htp]
	\centering
    \includegraphics[width=.85\textwidth]{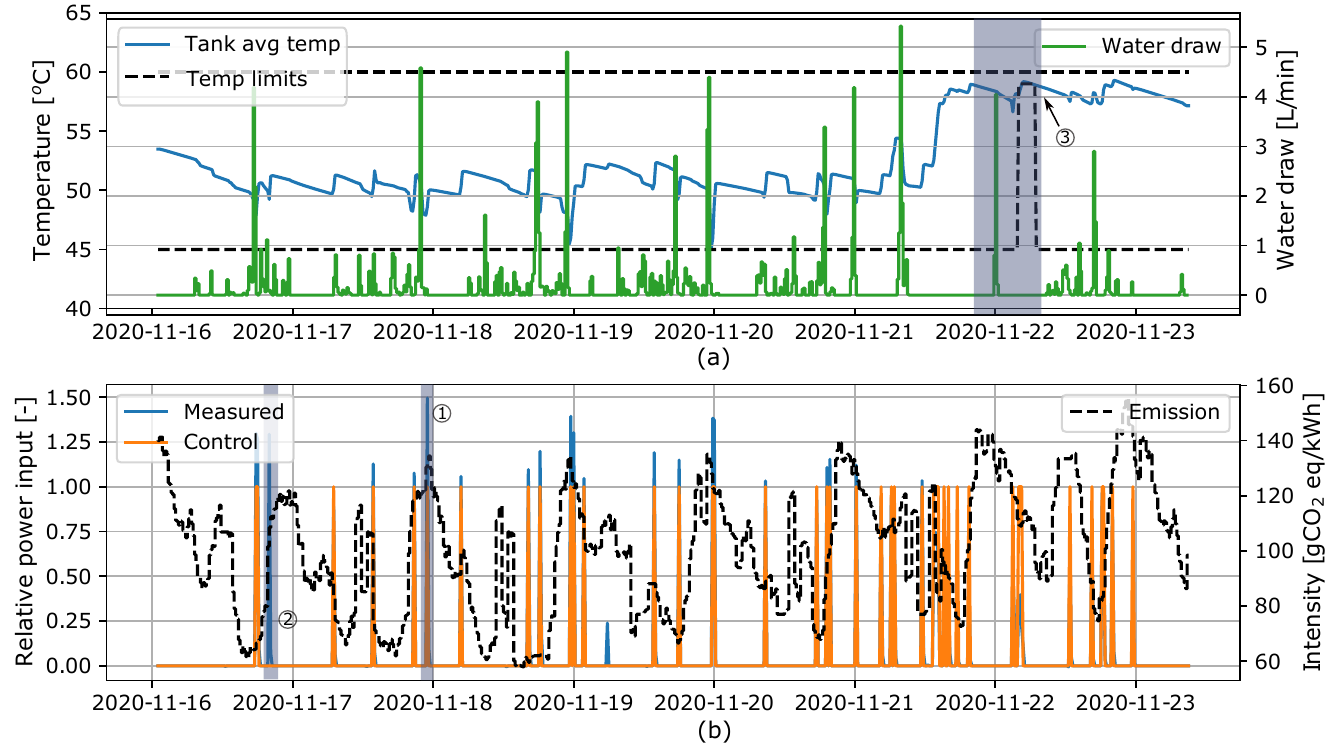}
	\captionof{figure}{Field experiment results of \ac{DHW}. (a) shows the average tank temperature and the water draw over time. The planned and realized thermal power input into the tank are summarized in (b), in which the thermal power is normalized with a thermal power capacity inferred from historical data. }\label{fig:exp_result_dhw_emission_nov}
\end{figure}

We can observe from Figure \ref{fig:exp_result_dhw_emission_nov}(a) that the temperature is always within the predefined limits, indicating adequate hot water supply. Right before the boost of the lower temperature limit, energy is actively used during the low carbon intensity period. 
However, \autoref{fig:exp_result_dhw_emission_nov}(b) reveals that power consumption doesn't always inversely correlate with carbon intensity. 
For instance, unexpected high-energy demands, possibly from substantial water usage, caused power surges during peak carbon intensity periods (marked by the grey period \circled{1} in \autoref{fig:exp_result_dhw_emission_nov}).
Additionally, the net power exchange with the grid shown in \autoref{fig:exp_result_grid_emission_nov} does not show a peak at this time (marked by the grey period \circled{1} in \autoref{fig:exp_result_grid_emission_nov}), implying that the power was supplied internally. We can also observe that the measured power input mostly follows control decisions, with exceptions such as that marked by the grey period \circled{2} in \autoref{fig:exp_result_dhw_emission_nov}, which is a result of actuation errors. Additionally, thermal power inputs are dependent on the difference between tank temperature and inlet temperature. Hence, the measured power input varies over time.

\subsubsection{Other assets and grid connection point}
\begin{figure*}[!htp]
	\centering
    \includegraphics[width=.85\textwidth]{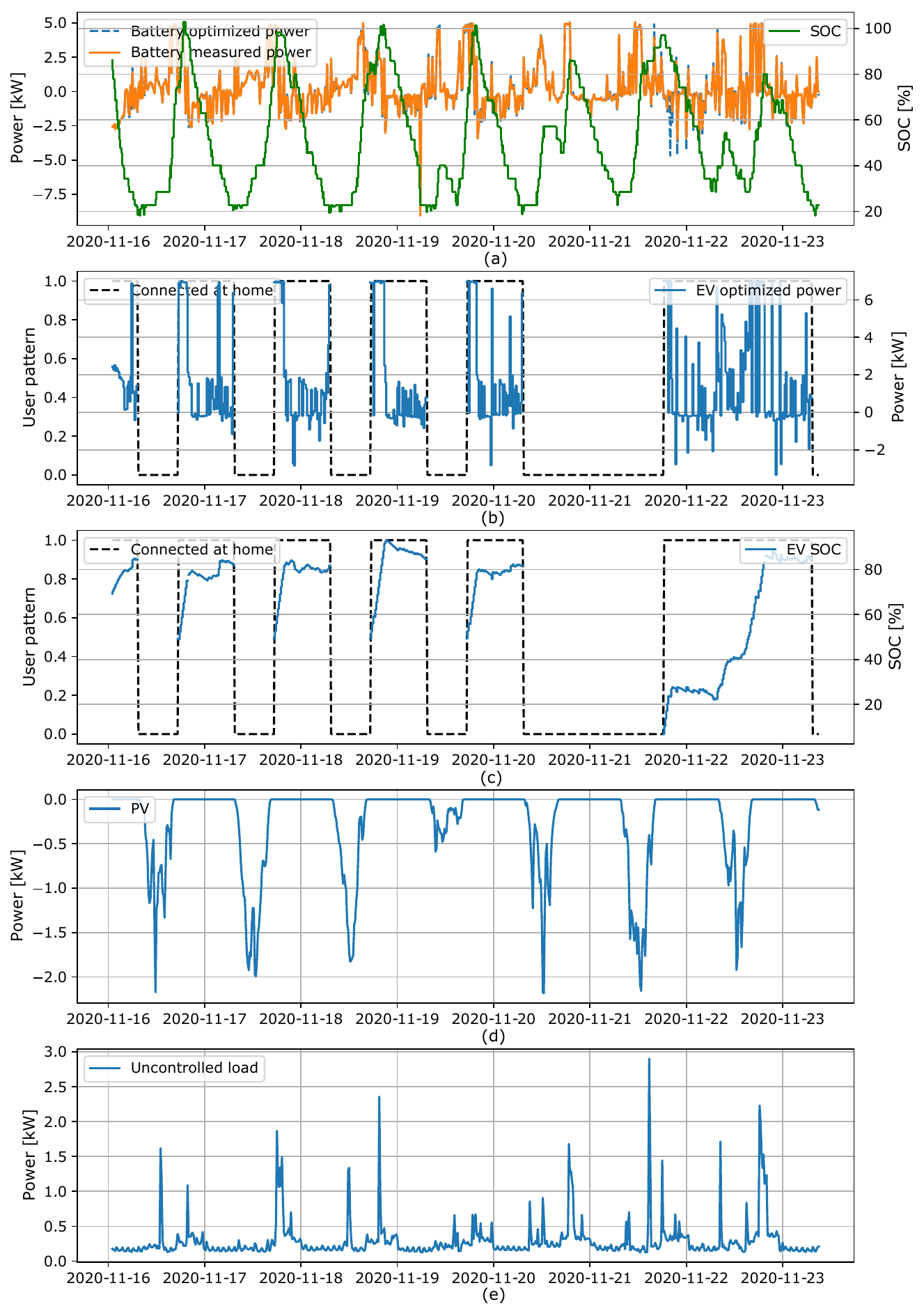}
	\captionof{figure}{Field experiment results of all non-thermal assets. (a) shows the battery's \ac{SOC}, planned/measured charging/discharging power. (b) shows the simulated \ac{EV}'s charging/discharging power. (c) shows the \ac{EV}'s \ac{SOC}. The dashed black line indicates the user's driving pattern, with 1 indicating that the \ac{EV} is connected to the charger at home and 0 indicating the opposite. (d) and (e) show the measured \ac{PV} power and the uncontrolled load respectively. }\label{fig:exp_result_nonthermal_emission_nov}
\end{figure*}

For brevity, the results of all non-thermal loads are summarized in \autoref{fig:exp_result_nonthermal_emission_nov}. Note that the bidirectional-charging \ac{EV} was absent; results in \autoref{fig:exp_result_nonthermal_emission_nov}(b) and (c) represent simulated \ac{SOC} and charging/discharging power.

\begin{figure*}[!htbp]
	\centering
    \includegraphics[width=.9\textwidth]{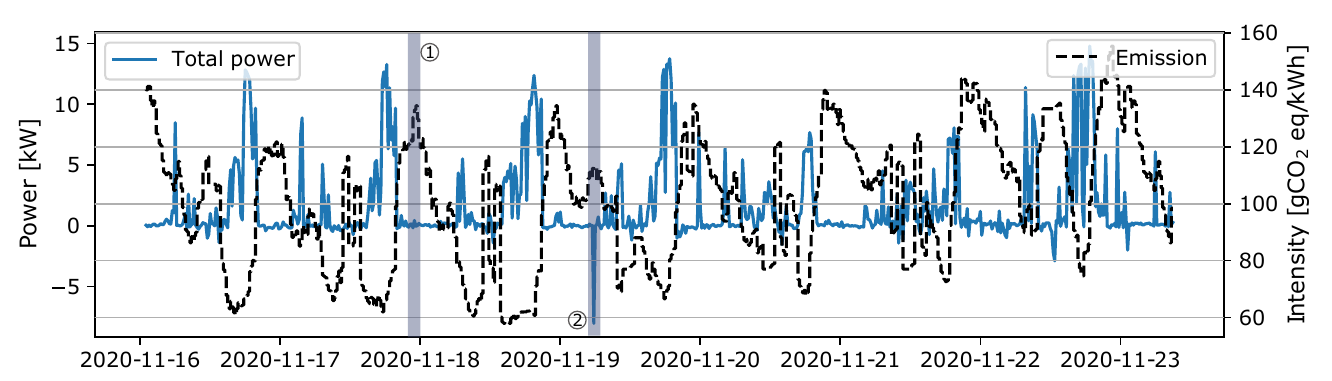}
	\captionof{figure}{Field experiment results of the net power exchange with the grid. The time series of the net electric power exchange is obtained by summing all respective power measurements and adding the simulated EV charging/discharging power.}\label{fig:exp_result_grid_emission_nov}
\end{figure*}

The net electric power exchange with the grid aggregating all the assets  is summarized in \autoref{fig:exp_result_grid_emission_nov} alongside the  electricity carbon intensity. 
Interestingly, the net power often inversely correlates with the carbon intensity profile. 
One notable grid export, during the grey period \circled{2}, can be traced back to an uncontrolled load profile's forecast error. 

\subsection{Impact evaluation}
To gauge the controller's efficacy in reducing emissions, this work contrasts the presented results with a simulation on a virtual testbed. Although a simplified model is used in the closed-loop control, it is not accurate enough as a virtual testbed to emulate the physical system over one week. Hence, the digital twin developed in \cite{khayatian2020temporal}, is used. This digital twin is calibrated using the data of entire 2019 at 1-minute intervals with a CV-RMSE of 0.09, making it a suitable choice for the virtual experiment.
Compared with other frequently used methods such as reference day, the proposed virtual experiment captures the high-resolution variation of carbon intensity profile. 

The virtual experiment employs a hysteresis controller for both \ac{SH} and \ac{DHW}, which is common in the current industry. 
More specifically, the average water tank temperature limits are [\SI{55}{\degreeCelsius}, \SI{60}{\degreeCelsius}] and the comfort zone of \ac{SH} is the same as the physical experiment. \ac{DHW} is treated differently to ensure a sufficient temperature level to eliminate Legionella, which also replicates the existing industry practice. The \ac{EV} is charged at full power right after being connected to the charger, and a hysteresis controller is applied afterward to ensure the minimum \ac{SOC} level. Currently, there are no standard industrial practices as to design energy management systems covering all flexible devices. A self-consumption-oriented strategy is considered for the battery in addition to the hysteresis controllers mentioned above. It stores all \ac{PV} production; for the rest of the time, it only discharges to cover the demand of other appliances. Note that the same carbon intensity profile is used in both the control and the impact evaluation. This assumes a perfect carbon intensity forecast. 
Interested readers are referred to \cite{aliunid_proj_report} for the forecast accuracy. 

\begin{figure*}[!htp]
	\centering
    \includegraphics[width=0.9\textwidth]{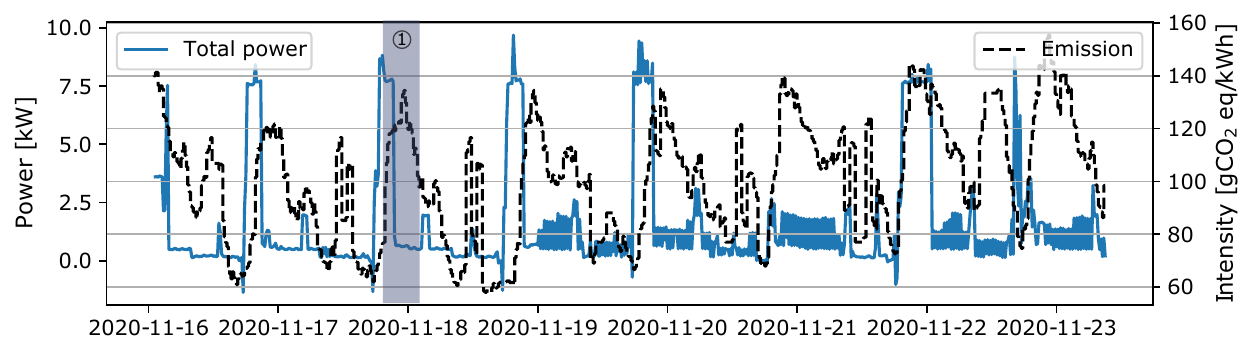}
	\captionof{figure}{Simulated results of the net power exchange with the grid.}\label{fig:exp_result_grid_emission_nov_bm}
\end{figure*}

We can observe that the total emission and the average carbon footprint of the consumed electricity are reduced by 12.5\% and 16.5\%, respectively. The reduction is realized by avoiding electricity imports during high carbon intensity periods. An example is marked by the grey period \circled{1} in \autoref{fig:exp_result_grid_emission_nov_bm} and \autoref{fig:exp_result_dhw_emission_nov}. The reduction is not pronounced, which indicates that the benchmark control strategy mentioned above already reduce carbon footprint by promoting self-consumption.

\subsection{Flexibility quantification and provision}
\begin{figure*}[!htp]
	\centering
    \includegraphics[width=0.8\textwidth]{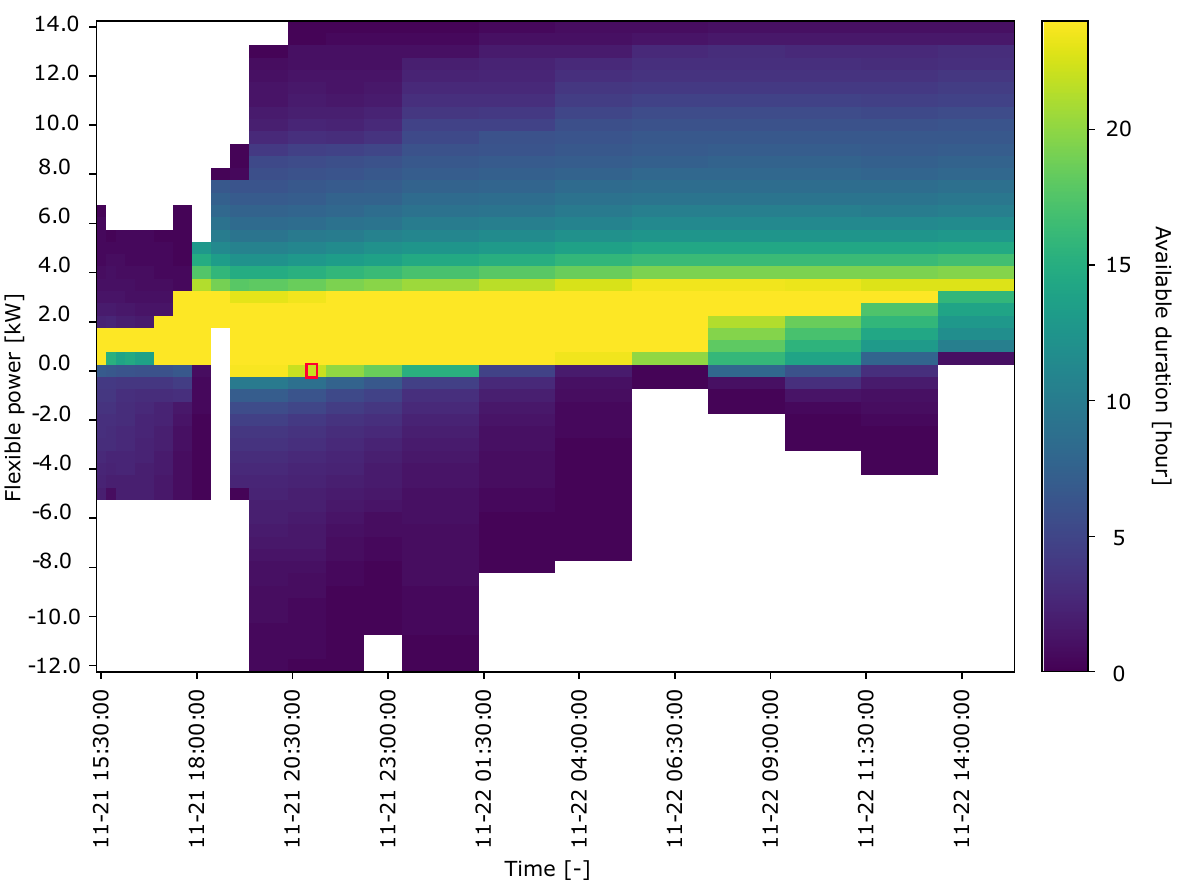}
	\captionof{figure}{An example of flexibility envelope exported by the controller.}\label{fig:flex_enve_202011211533_all_pv}
\end{figure*}

\begin{figure*}[!htp]
	\centering
    \includegraphics[width=.7\textwidth]{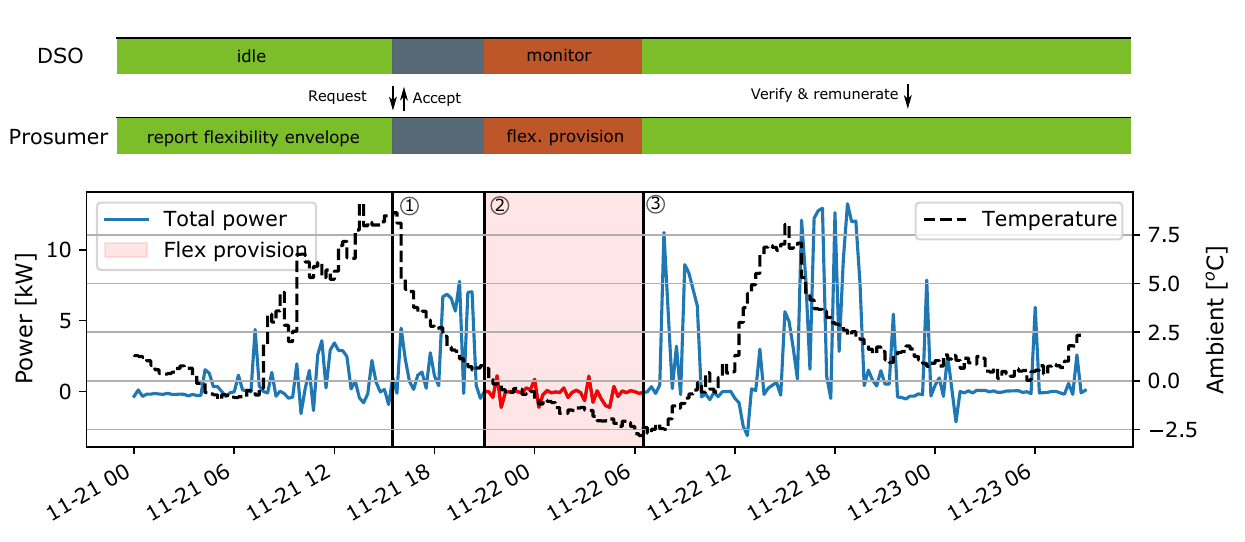}
	\captionof{figure}{Results of flexibility provision example. The bars on top of the figure denotes the actions between the emulated DSO and the prosumer. The blue and red curves denote the aggregate power of all flexibility resources outside and inside flexibility provision period respectively.}\label{fig:exp_result_grid_flex_emission_nov}
\end{figure*}

An example of online flexibility envelope quantification is provided in \autoref{fig:flex_enve_202011211533_all_pv}, which shows that flexibility potential considerably vary within the 24 hour horizon. 
Results of the flexibility provision with an emulated \ac{DSO} are in \autoref{fig:exp_result_grid_flex_emission_nov}.
We consider a scenario in which the \ac{DSO} experiences network congestion due to low ambient temperature. 
More specifically, load peaks may be exacerbated due to simultaneous consumption from newly adopted \acp{HP}, which the distribution system may not be planned for.
Thus, additional flexibility from buildings is needed to mitigate the issue. As per industry practice, ripple control \cite{westermann2007demand} has been used for decades for direct load control by broadcasting radio signals to cease operation of devices such as \acp{HP} within a target group. However, ripple control represents unidirectional communication and addresses limited types of flexible devices. 
In the rest of this section, we demonstrate the proposed framework with an emulated \ac{DSO}. 
More importantly, we show similar performance, namely keeping total power exchange with the grid close to 0 kW, can be achieved from the perspective of the \ac{DSO}, while comfort levels and preferences of end users are shown to be respected. 

In the experiment, a building self-reports its flexibility envelope to the \ac{DSO}, who in turn remains idle until flexibility needs are foreseen according to the weather forecast. The \ac{DSO} examines the reported flexibility envelope shown in \autoref{fig:flex_enve_202011211533_all_pv} and notifies flexibility provision to the building via  $(\tau_\text{s},\tau_\text{e}, P)$ := (2020-11-21 21:00:00+01:00, 2020-11-22 06:30:00+01:00, 0 kW) at the time marked by the vertical line \circled{1} in \autoref{fig:exp_result_grid_flex_emission_nov}. 
Importantly, $(\tau_\text{s},\tau_\text{e}, P)$ needs to match the self-reported flexibility envelope (marked by the red box in \autoref{fig:flex_enve_202011211533_all_pv}). Within the flexibility provision period (marked by the period between vertical lines \circled{2} and \circled{3}), the building tracks the set point. The results show that the total power exchange with the grid is reduced to a marginal level, although not strictly zero. This can be attributed to the actuation errors as observed in , \autoref{fig:exp_result_dhw_emission_nov} and \autoref{fig:exp_result_nonthermal_emission_nov}. Besides, we can observe that the energy states of all devices are comfortably away from their lower limits (as seen at the end of the grey area \circled{11} in the first plot of \autoref{fig:exp_result_sh_emission_nov}, and the end of the grey area \circled{3} in the first plot of \autoref{fig:exp_result_dhw_emission_nov}). This indicates that there are no immediate needs for electricity imports from the grid. Therefore, there is no risk of rebound effects. While the building activates its flexibility, the \ac{DSO} continuously monitors the building and remunerates the service provider afterwards. A full discussion of the remuneration scheme lies beyond the scope of this study, but this remains an important issue for future research.

\section{Discussion}\label{sec:discussion}
Overall, the presented results demonstrate that the controller operates with emission-aware \ac{MPC} as base strategy and can deviate from the optimal trajectory to provide flexibility upon request.
In the experiments, simplified \ac{SH} models, which were extracted from historical data, yielded  satisfactory results both in terms of emission reduction and maintaining thermal comfort. This might be attributed to the unit being heated with water-based ceiling panels, which have slower dynamics than forced air heating \cite{bacher2011identifying}. Additionally, the temperature is controlled within a small range. 

Apart from the quantitative data, qualitative feedback on thermal comfort was also collected via an online feedback form during the experiments \cite{exp_specs}. For \ac{DHW}, all feedback indicated ``very satisfied''. As for the bedrooms, 37.5\% of the time, the occupants indicated slightly cool indoor temperature in the 7-scale rating (cold, cool, slightly cool, neutral, slightly warm, warm, hot) matching low comfort violations. At other times, occupants expressed neutral opinions about the indoor temperature. We observed fatigue among users responding to survey requests, limiting the current survey. This suggests that the feedback strategy in the future needs to take a different form, especially for real-time control. Since the bi-directional \ac{EV} was emulated, no feedback was gathered.

Compared to traditional ripple control \cite{westermann2007demand}, the presented framework offers a significant advantage. It allows \acp{DSO} to obtain a comprehensive overview of available flexibility, effectively combining all flexibility resources. The high-resolution flexibility supports precisely resolve potential network issues. 
However, reliability might be a concern. This is because reliability decreases as the element count increases in any series system. The proposed prosumer is an example of such series systems. Additionally, ripple control is implemented in an open-loop fashion and the response can be expected within 7s \cite{westermann2007demand}, whereas the proposed framework would take longer to quantify flexibility envelope and establish flexibility provision agreements. Moreover, communicating the flexibility envelope requires significant bandwidth, necessitating further simplification. All in all, the existing ripple control scheme excels in simplicity and responsiveness. In contrast, the presented framework is favorable for \ac{DSO}s that require automation and an optimization-based approach due to the complexity of handling numerous resources.
\acresetall
\section{Conclusion}\label{sec:conclusion}
Despite buildings' promising active role in supporting the energy transition, challenges arise due to the involvement of diverse operational objectives and stakeholders. This work addresses gaps in experimental insights, focusing on emission-aware operation, flexibility quantification and provision to \ac{DSO}, and the impacts on/from occupants. 

During a week-long experiment, a 12.5\% reduction in equivalent emissions was achieved compared to a benchmark controller maximizing PV self-consumption. Meanwhile, measurements indicate that end users' comfort levels are improved. 
The proposed flexibility provision framework, demonstrated with an emulated distribution system operator, considers a scenario where flexibility is requested to mitigate network congestion.
All behind-the-meter flexible resources are effectively coordinated to provide flexibility upon notification. The experimental results suggest flexibility can be offered without rebound effects, while still maintaining user comfort levels and preferences.

Nonetheless, there are several limitations to note. Firstly, this work does not explicitly account for uncertainties associated with forecast and model errors in the optimization problem. The inclusion of uncertainties can enhance reliability. However, practical deployment costs must be balanced against this. Secondly,  challenges may arise if the maximum duration of flexibility provision is requested or when a \ac{DSO} requires flexibility with a different lead time. It remained to be assessed whether flexibility can be reliably provided and comfort guaranteed. Future research should assess more scenarios for robust flexibility quantification and provision. To better account for uncertainties, an alternative approach could be a probabilistic flexibility representation.

\section*{Acknowledgement}
The project was funded by the Swiss Federal Office of Energy (Section: Energy Research and Cleantech) under the project SI/501841 ``aliunid - Versorgung 'neu': Feldtest 1.1.2019 – 30.6.2020'' and the Sustainable Demand Side Management for the Operation of Buildings (S-DSM)
project under the contract number SI/502165-01. We would like to thank Reto Fricker, Sascha Stoller and Benjamin Huber for their support, and Fazel Khayatian for providing the digital twin of UMAR. We would also like to thank Charalampos Ziras, Julie Rousseau and Natasa Vulic for discussions and proofreading the draft. Finally, we acknowledge the members of IEA EBC Annex 82 for comments and suggestions.
\bibliographystyle{ieeetr}
\bibliography{reference} 

\end{document}